\documentclass[arxiv,usenatbib]{iupartex}

\usepackage{newtxtext,newtxmath}
\usepackage[T1]{fontenc}

\usepackage{graphicx}
\usepackage{blindtext}
\usepackage{amssymb}
\usepackage{multicol}
\usepackage[mathscr]{euscript}
\DeclareSymbolFont{rsfs}{U}{rsfs}{m}{n}
\DeclareSymbolFontAlphabet{\mathscrsfs}{rsfs}

\usepackage{array}
\usepackage{amsmath}
\usepackage{makecell}
\usepackage{caption}
\usepackage{subcaption}
\usepackage{float}
\usepackage{url}

\DeclareCaptionLabelFormat{subcaptionlabel}{\normalfont(\textbf{#2}\normalfont)}
\usepackage[normalem]{ulem}
\usepackage{xcolor}

\usepackage{lipsum}

\title[PAR LATEX template]{The First VHE Activity of OJ 287 and the Extragalactic Background Light}

\author[Yadav et. al 2024]{%
Sameer Yadav$^{1\cc}$,\orcid{0009-0007-2984-5882}
Pankaj Kushwaha$^{1}$,\orcid{0000-0001-6890-2236}
\affsep \\
$^1$Department of Physical Sciences, Indian Institute of Science Education and Research Mohali, SAS Nagar
 140306, India
}

\corres{S. Yadav; P. Kushwaha}{sameerydv574@gmail.com (S.Y.);  pankaj.kushwaha@iisermohali.ac.in (P.K.)}

\pubyear{2024}

\doiheader{XXXXXXX/PAR.20XX.00000}
\date{
	\pSubmit{00.00.0000} 
	\pRevReq{00.00.0000}
	\pLastRevRec{00.00.0000}
	\pAccept{00.00.0000}
	\pPubOnl{00.00.0000}
}
\volume{0}
\volnumber{1}

\begin{document}
\label{firstpage}
\pagerange{\pageref*{firstpage}--\pageref*{lastpage}}
\maketitle

\begin{abstract}
The BL Lacertae (BL Lac) object OJ 287 underwent an intense X-ray activity phase, exhibiting its brightest recorded X-ray flare in 2016-2017, characterized by much softer X-ray spectra and, concurrently,
its first-ever recorded very-high-energy (VHE) emission (100--560 GeV), reported by the VERITAS observatory. Broadband spectral energy distribution reveals a new jet emission component similar to high-synchrotron-peaked BL Lac objects, thereby implying the soft X-ray spectrum for the synchrotron emission. Using the advantage of simultaneous X-ray and VHE spectral information, as well as the source being a low-synchrotron-peaked BL Lac object, we systematically explored the extragalactic background light (EBL) spectrum by demanding that the VHE spectrum cannot be harder than the X-ray spectrum. We used three different phenomenological
forms of the EBL spectral shape (power-law, parabola, and polynomial) motivated by current constraints on the EBL with the Bayesian Monte Carlo approach to infer the credible EBL range.
Our study favors an almost flat power-law spectral shape and is consistent with previous studies. The other spectral forms capable of capturing curvature though result in a better statistics value; the improvement is statistically insignificant given the additional parameters.
\end{abstract}




\section{Introduction}

The Universe as apparent today has gone through many different phases of evolution 
since beginning from a very hot and dense phase---the Big Bang. Each of these different
evolutionary phases is expected to be associated with a distinct radiation signature, e.g.,
the cosmic microwave background (CMB) is the relic associated with the decoupling of radiation--matter
in the matter-dominated phase ($\rm z$$\sim$1100). One of the milestones in this evolutionary sequence
is gravity locally becoming dominant, marking the start of structure formation, i.e.,~the first
stars with copious optical and ultraviolet emission. A~significant part of this radiation is expected
to be re-processed to infra-red energies through interactions with matter/dust \citep{Primack_2005}. The~extragalactic background light (EBL) is this cumulative
radiation in the wavelength range of 0.1--1000 $\upmu$m \citep{Dwek_2013} produced through this cosmic
history, from~both resolved and unresolved sources. The~inferred spectral shape of the EBL is bimodal with
a peak at around 1 $\upmu$m, attributed to be from direct starlight emission (but see \citep{Lauer_2022})
and the other peak at around 100 $\upmu$m argued to be from re-emission of absorbed starlight by the dust
that encodes the formation and evolution history of galaxies and the structure formation of the~Universe.

\begin{figure}[H]
    \includegraphics[scale=0.54]{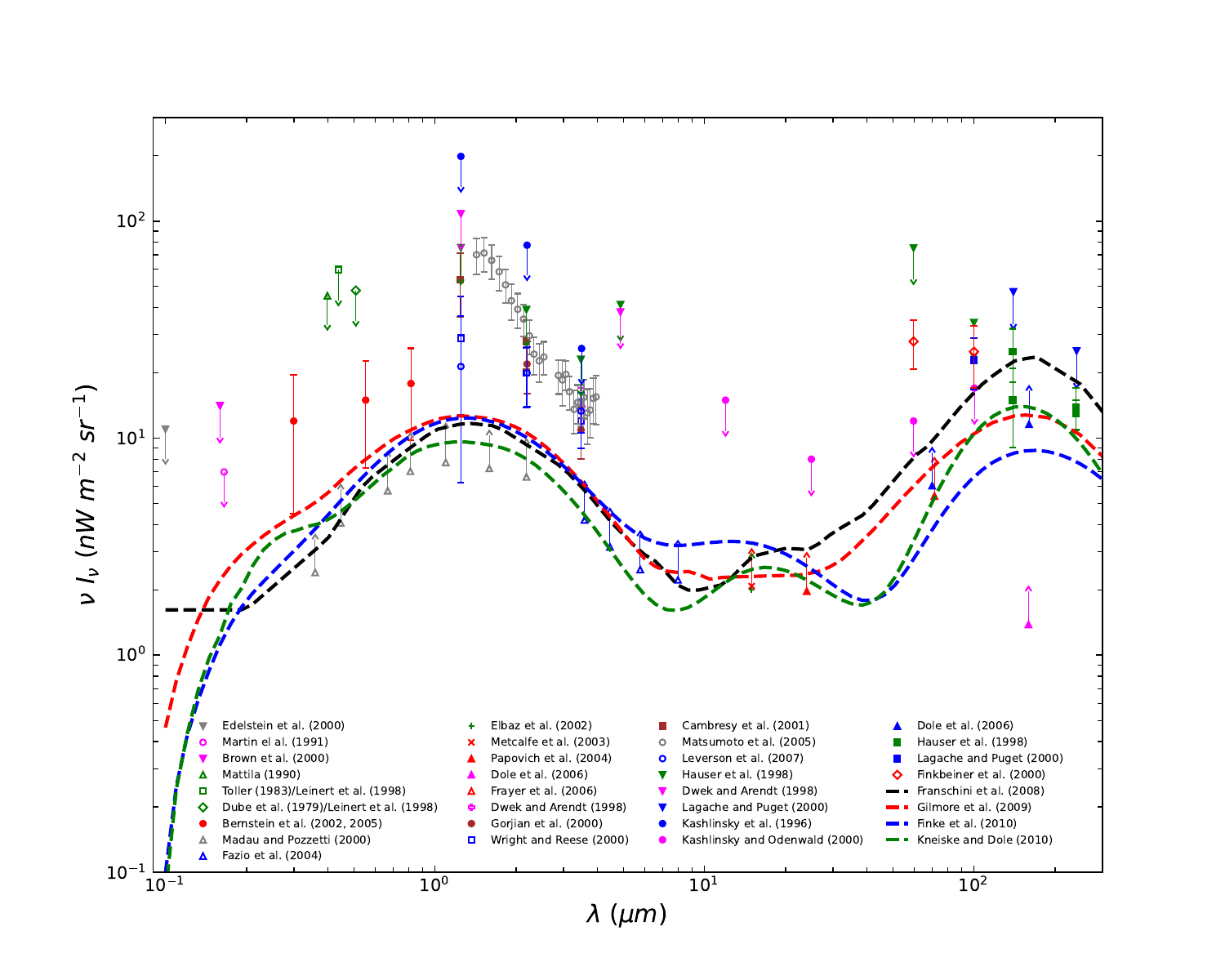} 
    \caption{EBL SED from direct and indirect measurement. Data points are from direct measurements \citep{Mazin_2007}, and the curves are the theoretical models\protect\footnotemark ~predicting limits on the~EBL.}
    \label{fig:Fig. 1}
\end{figure}

Direct measurement of the EBL is a very challenging task, primarily due to contamination by the Zodiacal and galactic light \citep{Hauser_2001, Lauer_2022}. However, lower limits on the EBL intensity have been provided by the integrated light from resolved galaxies \citep{Madau, Fazio, Frayer_2006,Dole_2006}, and upper limits have been derived by direct observations of the EBL with space- and ground-based instruments sensitive in specific bands \citep{Hauser_1998,Kashlinsky_1996,Gispert_2000}. Most of the EBL is produced by discrete galactic and primordial stellar sources. Studying the fluctuations in their number and clustering properties has provided the spatial fluctuations in the EBL intensity that provide a limit on its intensity through fluctuation analysis 
 \citep{Kashlinsky_2000}. These limits indicate a bimodal structure of the EBL between the UV and far-IR with a peak at $\rm\sim$1 $\upmu$m and the other at $\rm \sim$200 $\upmu$m, but the absolute spectral energy distribution (SED) of the EBL remains uncertain and highly debatable. Several theoretical models predicting the SED of the EBL at $z = 0$ have been developed (see Figure~\ref{fig:Fig. 1}; \citet{Franceschini_2008,Gilmore_2009,Finke_2010,Kneiske_2010}), but the lack of observations between the two peaks and the observational challenges have impeded its measurement. Backward-evolution models start with the determination of the luminosity density, $\mathscrsfs{L}(\lambda, z=0)$, in~the local Universe and then evolve it with redshift at different wavelengths \citep{Franceschini_2008, Dominguez_2011}. Forward-evolution models use the cosmic star formation rate (CSFR) to calculate $\mathscrsfs{L}(\lambda, z=0)$ determining the distribution of the energy over wavelengths \citep{Finke_2010}. Semi-analytic models calculate $\mathscrsfs{L}(\lambda, z=0)$ by taking physical processes into account for the formation and evolution of the structure of 
the Universe \citep{Gilmore_2009}.

A complimentary and powerful probe to put limits on the EBL is by analyzing the signatures of the EBL on the very-high-energy (VHE; E $>$ 100 GeV) gamma-ray spectrum of distant blazars \citep{Georganopoulos_2010, Ackermann_2012, HESS_2013}, provided the intrinsic VHE spectrum of the source is known. The~VHE gamma-rays while traversing
the intergalactic medium interact with the EBL photons and get
absorbed by the process of electron--positron pair production \citep{Gould_1967}. The~absorption of VHE photons by the EBL steepens the VHE spectra, hence providing the imprints of the EBL (e.g., \citep{Vassiliev_2000, Mankuzhiyil_2010}). Dis-entangling the EBL-induced attenuation signature in the observed spectrum and the intrinsic spectrum offers a powerful tracer of EBL density. Thanks to the tremendous overall advancement in observational gamma-ray astronomy, as well as blazar's SED modeling, several studies have attempted to constrain the EBL based on the attenuation of gamma-rays (e.g., \citep{Stecker_1993,Hauser_2001,Dwek_2005,Mazin_2007}). For example, using the H 1426 + 428 spectrum, along with the spectra of already discovered TeV blazars, limits on the EBL were derived \citep{Costamante_2004,Kneiske_2004}. Similarly, HESS observations of 1ES 1101-232 and H 2356-309 were used to estimate the EBL in optical--NIR wavelengths by assuming the intrinsic spectrum cannot be harder than $\Gamma_{int}$ = 1.5 \citep{Aharonian_2006}. However, subsequent EBL studies 
argued for a much harder particle spectrum of distant blazars \citep{2008ApJ...689L..93K}, and now, many sources have shown a very hard photon spectrum, implying a harder particle spectrum ($\Gamma_{int}$$\sim$$1$), and magnetic re-connection has been argued to be the most prominent driver. Contrary to this,
studies have also demonstrated that, in certain scenarios, the~hardness can be well reproduced, e.g.,~internal gamma-ray absorption through interaction
with narrow-band strong photon fields \citep{2008MNRAS.387.1206A,2011ApJ...738..157Z} without requiring a hard particle~spectrum.

The first systematic exploration of the EBL signature in blazar spectra at VHEs was presented in \citet{HESS_2013}, reporting the EBL signature at the 
$8.8\sigma$ level with respect to the model of \citet{Franceschini_2008} and constraining it in $0.3~\upmu \text m < \lambda < 17 ~\upmu \text m$. This approach was
similar to the one introduced in the Fermi large area telescope (Fermi-LAT) collaboration EBL work \citep{Ackermann_2012}. The~Fermi-LAT team
subsequently expanded this work, exploiting improved exposure data of 739 blazars and found attenuation due to the EBL in the entire explored redshift
range $0.03 < z <3.1$ for all twelve models that were \mbox{tested \citep{Fermi_2018}}. Using this, they also reconstructed EBL evolution, as well as
the cosmic star formation history with the latter being consistent with other independent studies. It should, however, be noted that, for sources
exhibiting strong spectral changes, using long-exposure data may introduce false EBL~signatures. 

Blazars currently have been divided into three primary spectral classes: low-\linebreak synchrotron-peaked (LSP), 
intermediate-synchrotron-peaked (ISP), and~high-synchrotron-peaked (HSP), based on 
the band/frequency in which the low-energy component of their bi-model SED attains the  
maximum \citep{Abdo2010}. This scheme is an extension of the previous existing
classification designating BL Lacertae (BL Lac) objects as low-peaked (LBLs), intermediate-peaked (IBLs), and~
high-peaked (HBLs) with LSP now representing 
FSRQs and LBLs. The~majority of the blazars that have exhibited frequent VHE 
activities belong to the HBL/HSP class and are strong X-ray emitters. The~X-ray (or
low-energy component of the SED) is understood to be synchrotron, while high-energy is 
primarily inverse Comptonization (IC) (e.g., \citep[][]{Bottcher2013, 2019NatAs...3...88G}). A~simultaneous 
X-ray and VHE spectral measurement can be used to study the EBL, except that the HBL VHE
can be significantly steepened due to the Klein--Nishina effect, thereby overestimating the EBL \citep{2008ApJ...679L...9B}. Such systematic exploration
, however, is expected to not affect the
outcome when using a sample of blazars at different redshifts. LBL's gamma-rays, on~the other hand, are primarily due to the IC of infra-red photons and are relatively less prone to the IC effect, except that only a few counted ones are known\footnote{TeVCat: online catalog for TeV astronomy; see \url{http://tevcat.uchicago.edu} accessed on 22nd November 2023.
}, and thus, at~best, this can provide
a constraint on the limited range. Nonetheless, this can be used to confirm and cross-
check the current EBL~limits.

OJ 287 is a BL Lac-type object  located at z = 0.306 \citep{Sitko_1985} and is one
of the three BL Lac sources detected at VHEs \citep{obrien2017veritas}. BL Lacs are jetted active galactic nuclei (AGN) with a very weak or a complete absence of emission line features in their optical spectrum 
 [equivalent width $<5~$A] and are generally accepted to lack the standard accretion disk and infra-red torus. 
They emit an entirely jet-dominated highly variable continuum covering the entire accessible electromagnetic spectrum from radio to GeV--TeV gamma-rays. The~broadband emission exhibits a characteristic broad dual-humped spectral energy distribution \citep[][]{Abdo2010}. In~the broadband SED-based categorization, OJ 287 is classified as a low-energy-peaked BL lac (LBL/LSP) object, and its gamma-ray emission is primarily attributed to the inverse Comptonization of the IR photon field \citep{Kushwaha_2013}.
The inverse Compton interpretation of MeV--GeV gamma-rays is also strongly favored by the SED 
modeling of candidate neutrino blazars, indicating a sub-dominant hadronic contribution (e.g., \citep[][]{2019NatAs...3...88G}).

OJ 287 exhibited its first-ever very-high-energy (VHE; E $>$ 100 GeV) $\gamma$-ray activity in 2016--2017, reported by the Very Energetic Radiation Imaging Telescope Array System (VERITAS), which continued for a very long time \citep{obrien2017veritas}. The broadband study of the VHE period along with previous and later activity episodes
revealed an additional HBL-like emission component, establishing the soft X-ray emission to synchrotron \citep{Kushwaha2018b}, as is the case in HBLs. This VHE emission can
be explained via the HBL/HSP scenario or the IC of an IR photon field. Given that gamma-ray
emission of LBLs/LSPs is due to the IC-IR \citep{Kushwaha_2013, Arsioli_2018}, we believe that the
VHE is also likely of the same~origin.

\addtocounter{footnote}{-1}

\footnotetext{ebltable: Python package for EBL models; see \url{https://github.com/me-manu/ebltable.git} accessed on 22 November 2023.}

In this work, we have carried out a systematic investigation of the reported VHE spectrum of OJ 287 to explore the EBL imprints it carries and, thereby, infer the EBL. In~\mbox{Section \ref{sec:PairProd}}, we revisit the basics of the photon--photon pair production mechanism 
and the relation between the intrinsic and observed spectrum. In~Section \ref{sec:obsConstraint}, we describe the constraints from the observations on the VHE spectral shape imposed by the general physical principles of the blazar's radiative processes. The~methodology/formalism to constrain the free parameters and the selection criteria of the models is explained in Section \ref{Method}. The~results are discussed in Section \ref{Results}. A~flat $\Lambda$ cold dark matter ($\Lambda$CDM) cosmology is used in this work with Hubble constant $H_{0} = 70~\text{km~s}^{-1}~\text{Mpc}^{-1}$; matter density parameter $\Omega_{m}$ = 0.27; and curvature density parameter $\Omega_{\Lambda}$ = 0.73.

\section{Attenuation of Gamma-Ray Photons by Pair~Production}\label{sec:PairProd}

The VHE $\gamma$-rays traversing the Universe are attenuated via pair production through interaction with low-energy EBL photons, $\gamma_{VHE} + \gamma_{EBL} \rightarrow e^{-} + e^{+}$, and the~following condition must be satisfied by the energies of two photons for the creation of the electron--positron pair:
\begin{align}
\label{eqn:threshold}
    E_{\gamma} \epsilon_{th} (1-\mu)=2m_{e}^{2} c^{4}
\end{align}
where $E_{\gamma}$ and $\epsilon_{th}$ are the energies of VHE $\gamma$-rays and the threshold energy for EBL photons, respectively, and~$\mu = \cos\phi$, such that $\phi$ is the angle between the momenta of two photons and $m_{e}$ is the rest mass of an~electron.

The cross-section for the $\gamma-\gamma$ interaction \citep{Gould_1967} is given by
\begin{align}\label{cross-section}
    \sigma_{\gamma\gamma}(E_{\gamma}, \epsilon, \mu, z) = \frac{3\sigma_{T}}{16}(1 - \beta^{2})\left[2\beta(\beta^{2} - 2) + (3 - \beta^{4})\ln \left(\frac{1 + \beta}{1 - \beta}\right) \right]
\end{align}    
where $\sigma_{T}$ is the Thompson cross-section and $\beta$ is given by
\begin{align}
    \beta = \left( 1 - \frac{2m_{e}^{2}c^{4}}{E_{\gamma}\epsilon(1-\mu)} \right)^{1/2} = \sqrt{1 - \frac{\epsilon_{th}}{\epsilon}}
\end{align}

The optical depth for pair creation of a $\gamma$-ray photon emitted by a source at a redshift $z$ via interacting with the EBL field \citep{Dwek_2013} is given by
\begin{align}
    \tau_{\gamma\gamma}(E_{\gamma}, z) = \int_{0}^z dz^{'} \frac{d\textit{l}}{dz^{'}} \int_{-1}^{1} d\mu \frac{1 - \mu}{2} \int_{\epsilon_{th}^{'}}^{\infty} d\epsilon \textit{n}_{\epsilon}(\epsilon, z^{'}) (1 + z^{'})^{3} \sigma_{\gamma\gamma}(\beta^{'}, z^{'})
\end{align}    
where \textit{l} is the proper distance such that
\begin{align}
    \frac{d\textit{l}}{dz} = c\frac{d\textit{t}}{dz} = \frac{c}{H_{0}(1 + z)[\Omega_{m}(1 + z)^{3} + \Omega_{\Lambda}]^{1/2}}
\end{align}
assuming a flat $(\Omega_{K} = 0)$ and matter-dominated $(\Omega_{R} << 1)$ $\Lambda$CDM Universe. $\textit{n}_{\epsilon}(\epsilon, z) = \frac{d\textit{n}(\epsilon, z)}{d\epsilon}$ is the specific comoving number density of the EBL.~$(1 + z)^{3}$ is the conversion term from the specific to the proper number density. 
In this context, the~threshold condition leading to pair production is modified to
\begin{align}
\label{eqn:cosmoThreshold}
    \epsilon_{th}^{'} = \frac{2(m_{e}c^2)^{2}}{E_{\gamma}\epsilon(1 - \mu)(1 + z)}
\end{align}
with $\beta$ becoming $\beta^{'} = \sqrt{1 - \frac{\epsilon_{th}^{'}}{\epsilon}}$.

Calculating the pair opacity of $\gamma$-rays requires the knowledge of the EBL comoving specific photon number density ${\textit{n}}_{\epsilon}(\epsilon, z)$ as a function of redshift. The~conversion of specific comoving intensity to comoving specific photon number density is given by
\begin{align}
    \epsilon^{2}\textit{n}_{\epsilon}(\epsilon, z) = \frac{4\pi}{c}\nu I_{\nu}(\nu, z) = 2.62\times10^{-4}\nu I_{\nu}(\nu, z)
\end{align}
where $\epsilon$ is the energy of EBL photons and the coefficient in the second term is calculated for $\epsilon$ in eV, $\textit{n}_{\epsilon}$ in $\text{cm}^{-3}\text{eV}^{-1}$, and~$\nu I_{\nu}$ in $\text{nW m}^{-2} \text{sr}^{-1}$.

The pair opacity leads to the steepening of the observed VHE spectrum from the intrinsic VHE $\gamma$-ray spectrum $\left( \frac{dN_{\gamma}}{dE} \right)_{int}$ emitted from the source due to EBL absorption, and the observed spectrum $\left( \frac{dN_{\gamma}}{dE} \right)_{obs}$ is related to intrinsic one as:
\begin{align}
    \left( \frac{dN_{\gamma}}{dE} \right)_{obs} = \left( \frac{dN_{\gamma}}{dE} \right)_{int} e^{-\tau(E_\gamma, z)}
\end{align}    

Thus, in~this approach, knowledge of the intrinsic spectrum is required to estimate the EBL. It should be noted
that the kinematics of the process and the energy dependency of the pair cross-section ($\sigma_{\gamma\gamma}$; 
Equation~(\ref{cross-section})) can change the observed VHE spectrum from a simple power-law to more complex with steepening
followed by a hardening depending on $E_{\gamma}$, and thus, a correct understanding of the EBL is essential to explore the VHE
and TeV emission mechanisms (e.g., \citep[][]{2013APh....43..241M}).

\section{Observational~Constraints}\label{sec:obsConstraint}

VERITAS detected an almost persistent, marginally variable VHE $\gamma$-ray emission from OJ 287 over a relatively extended duration from December 2016--March 2017 at $\sim$$10\sigma$ above the background in the energy range 100 GeV--560 GeV \citep{obrien2017veritas}. The~total time-averaged VHE spectrum is consistent with a power-law photon spectral index ($\Gamma_{VHE}$) of 3.49 $\pm$ 0.28. The~total time-averaged flux above 150 GeV for the entire dataset is $(4.61 \pm 0.62)\times 10^{-12} \text{cm}^{-2}\text{s}^{-1}$, which corresponds to $\sim$$(1.3 \pm 0.2)\%$ of the flux of the Crab Nebula above the same \mbox{threshold \citep{Hillas1998}}.

The reported VHE activity followed in the declining phase, after~the brightest-ever-recorded X-ray flare of the source \citep{Kushwaha2018b, komossa_2020}. Detailed studies and investigations of the spectral and temporal behavior establish it as a jet emission 
component as well, but of a different spectral class (HBL-like) \citep{Kushwaha2018b, 2022JApA...43...79K}  than the traditional spectral class of the source~\cite{Abdo2010, Kushwaha_2013}. Such an appearance of a new spectral class is very rare in~blazars.

As stated, for~OJ 287, an~external photon field at a temperature of $250K$ as soft target photons up-scattered by the highly relativistic electrons present in the 
jet by the IC mechanism has been argued to explain the high energy part of the spectrum 
\citep{Kushwaha_2013} in the LBL spectral state of the source while the 2015 gamma-ray 
spectral change was explained by the IC-BLR \citep{Kushwaha2018a, 2020Galax...8...15K}. The~photons emitted from the broad-line region (BLR) and the infrared (IR) emission from the dusty torus have typical photon energies $\epsilon_{BLR} \simeq 10~\text{eV}$ and $\epsilon_{IR} \simeq 0.3~\text{eV}$ ($\sim 1200~\text{K}$), respectively. For~the IC scattering of soft photons $\epsilon_{soft}$, the Lorentz factor of ultra-relativistic electrons that can produce the VHE emission is $\gamma_{e,IC} \simeq [(1+z)\epsilon_{VHE,obs}/(\delta_{source}\epsilon_{soft})]^{1/2}$, where $\delta_{source}$ is the Doppler factor and $\epsilon_{VHE,obs}$ is the typical VHE photon energy taken to be 100 GeV for our purposes. The~scattering occurs in the Thomson regime if $\gamma_{e,IC}\epsilon_{soft}/(m_{e}c^{2}) < 1$, which modifies to
\begin{align}
    \epsilon_{soft} < \frac{\delta_{source}(m_{e}c^{2})^{2}}{(1+z)\epsilon_{VHE,obs}} .
\end{align}

If the emission region is located within the external field extent, the~energy density of the external field in the comoving frame is boosted by a factor of $\Gamma_{source}^{2}$ and the photon energy increases by a factor of $\Gamma_{source}$, where $\Gamma_{source}$ is the bulk Lorentz factor. Hence, the~Thomson limit is $\epsilon_{ER} < 1.9 \text{eV} \times (\delta_{source}/\Gamma_{source})$. For~our study, we have taken bulk Lorentz factor $\Gamma_{source} = 16.5 \pm 4.0$ and Doppler factor $\delta_{source} = 18.9 \pm 6.4$ computed by \citet{Jorstad_2005} from 17 epochs at a 7 mm wavelength with the Very Long Baseline Array (VLBA) from 1998--2001. As~a result, only IR photons are scattered in the Thomson~regime.

\begin{figure}[H]
    \begin{subfigure}{.5\textwidth}
        \includegraphics[width= \linewidth, height= 6cm]{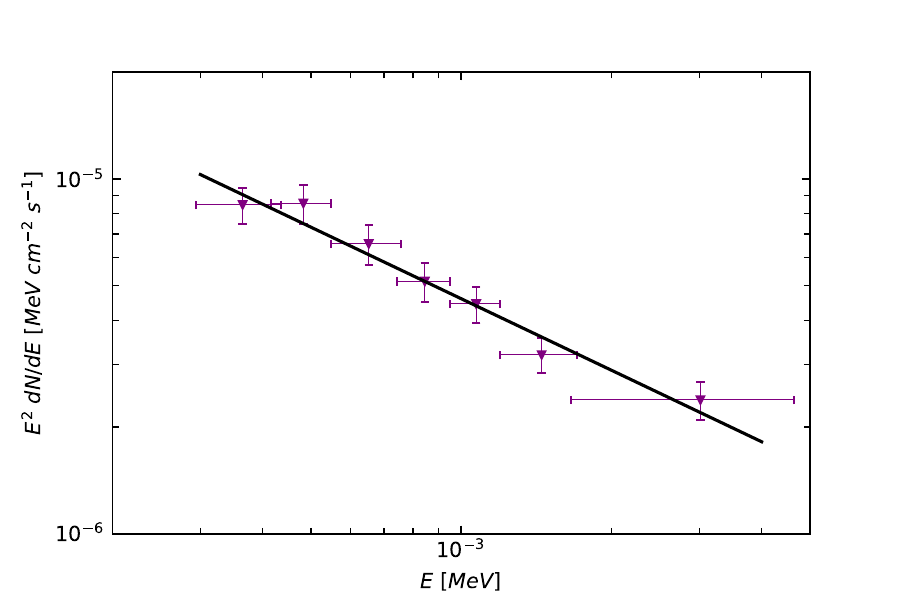}
    \end{subfigure}%
    \begin{subfigure}{.5\textwidth}
        \includegraphics[width= 0.95\linewidth, height= 5.4cm]{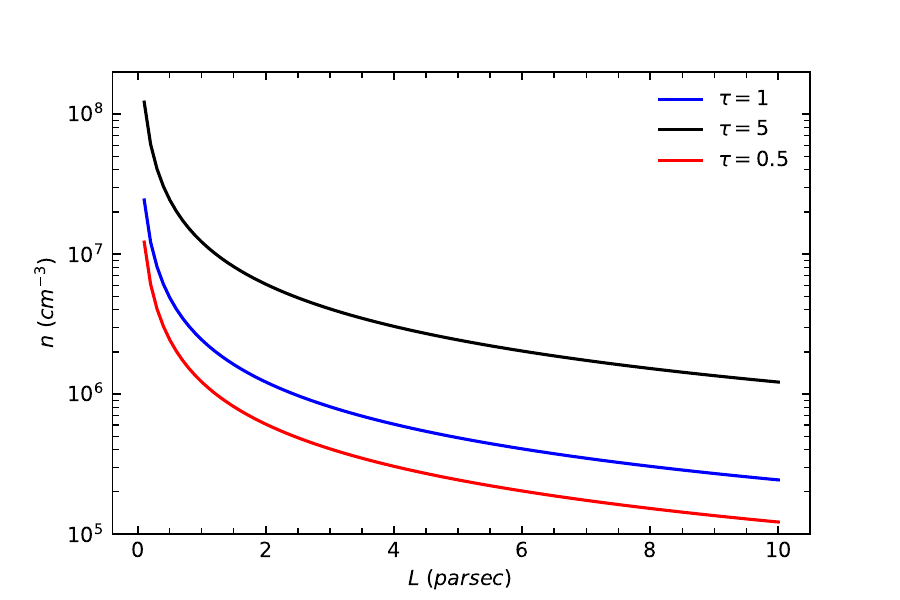}
    \end{subfigure}%
    \caption{\textbf{Left}: {\em Swift-XRT} measurements of OJ 287. The~line is the best fit of a power-law ($\Gamma = 2.66 \pm 0.05$). \textbf{Right}: Photon number density ($n$) within the BLR corresponding to different optical depths ($\tau$) for 100 GeV~photons.}
    \label{fig:Fig. 2}
\end{figure}

The external field up-scattered to gamma-rays via the IC can, in turn, also interact with the
emitted VHE photons. The~VHE gamma-ray photons upon interaction with the ER field are destroyed in the photon--photon pair production process when 
the condition in Equation~(\ref{eqn:threshold}) is satisfied. The~BLR radiation can absorb VHE photons of observed energy:
\begin{align}
    \epsilon_{VHE,BLR,obs} \gtrsim \frac{(m_{e}c^{2})^{2}}{(1+z)\epsilon_{BLR}} \simeq 20 \left(\frac{\epsilon_{BLR}}{10.2 eV}\right)^{-1} \text{GeV} 
\end{align}
and for IR radiation, the~observed threshold is
\begin{align}
    \epsilon_{VHE,IR,obs} \gtrsim \frac{(m_{e}c^{2})^{2}}{(1+z)\epsilon_{IR}} \simeq 665 \left(\frac{\epsilon_{IR}}{0.3 eV}\right)^{-1} \text{GeV} 
\end{align}
which is beyond the VHE energies detected by VERITAS. The observation of the VHE instead
indicates that the pair opacity is not high within the source. A~plot of the BLR photon number
density for different opacity values is shown in Figure~\ref{fig:Fig. 2} (right) for a 100 GeV~photon.

The above gamma-ray opacity arguments imply that the VHE emission is solely due to the IC scattering of IR photons from the dusty torus. This is also consistent
with the report of the association of the VHE activity with a new radio feature at parsec scales ($\rm\sim$10 pc,~\cite{2022A&A...658L..10L})---the putative location of the IR torus in the standard radio-loud AGN
unification scheme. Detailed and systematic investigations of broadband SEDs during the VHE phase
compared to the well-known OJ 287 SED (LBL/LSP; e.g.,~\cite{2022JApA...43...79K}) show a hardening of the optical--UV spectrum and a much softer X-ray,
implying a peak at UV energies (e.g.,~\cite{Kushwaha2018b, 2022MNRAS.509.2696S, 2020Galax...8...15K, 2022JApA...43...79K, obrien2017veritas}). A~similar hardening was observed at MeV--GeV energies, accompanied by the reported softer VHE spectrum,
implying an additional peak at GeV energies. A~comparison with the well-known blazar spectral sequence~\cite{Abdo2010} establishes that such spectral
features are characteristics of HBL/HSP emission as stated above and argued in previous works as well (e.g.,~\cite{Kushwaha2018b, 2022MNRAS.509.2696S}).

Since the IC-IR occurs in the Thomson regime, in~the standard blazar emission paradigm, the~intrinsic VHE
spectrum should be the same as the corresponding synchrotron
spectrum in the case of both originating from the same particle population, i.e.,~from one emission region, as~is the case here with the X-ray and VHE associated with the new HBL-like component. Thus,
the VHE spectral index ($\Gamma_{VHE}$) cannot be harder than the corresponding synchrotron counterpart, i.e.,~the $X$-ray 
spectral index here ($\Gamma_{X-ray}$). A~fit of a power-law $\frac{dN}{dE} = N_{0}(\frac{E}{E_{0}})^{-\Gamma}$ to \textit{Swift}-XRT data (Figure \ref{fig:Fig. 2}-left) results in 
a satisfactory fit ($\chi^{2}/d.o.f = 3.15/5$) with flux normalization $N_{0}= (3.32 \pm 1.02)\times10^{-6}~\text{MeV}^{-1}\text{cm}^{-2}\text{s}^{-1}$ and spectral index $\Gamma_{X-ray}= 2.66 \pm 0.05$ at normalization energy $E_{0}=0.2~\text{TeV}$, providing the maximum hardness the intrinsic 
VHE spectrum can attain, $\Gamma_{VHE}^{max} \simeq \Gamma_{X-ray} = 2.66$. It should be 
noted that the majority of the X-ray spectra during this VHE activity duration and such 
extremely soft X-ray state have been shown to be log-normal \citep{Kushwaha2018b,Kushwaha_2021,2022JApA...43...79K}, rather than a simple power-law. 
However, this log-normal part appears at the high-energy end of the X-ray band and is due 
to the traditional LBL spectral state of the source \citep{2022MNRAS.509.2696S}. The~NuSTAR observation during an intermediate spectral state also supports a simple
power-law spectrum up to \linebreak 30 keV. Beyond~this, the~spectrum is background-dominated 
\citep{2022MNRAS.509.2696S}. Thus, within~the current context and practical limitations, the~extremely soft X-ray spectrum associated with the VHE activity is
consistent with a~power-law.

The range of TeV detection limits the range over which the EBL can be explored/constrained. This can be seen from the threshold condition (Equation \eqref{eqn:threshold} or \eqref{eqn:cosmoThreshold}). For~a given VHE energy range, the~
corresponding wavelength range that can be probed is given by
\begin{align}
    \lambda_{ebl,max} = \frac{h}{2 m_{e}c}\times\frac{E_{VHE}}{m_{e}c^{2}}\times(1 + z)^{2} ~ \sim ~ 95~\upmu \text{m} \times \left(\frac{E_{VHE}}{20~\text{TeV}}\right) \times (1 + z)^{2}.
\end{align}

For OJ 287, the~observed VHE range of 100 GeV to 560 GeV corresponds to an EBL wavelength range of  0.8--2~$\upmu$m.

\unskip

\section{Analysis~Method}\label{Method}

With the above restriction on the VHE intrinsic spectrum, we performed the maximum likelihood analysis using the Markov chain Monte Carlo (MCMC) algorithm to constrain the EBL flux density. The~MCMC code \textbf{emcee} \citep{Foreman-Mackey_2013}, a~Python implementation of an affine invariant MCMC ensemble sampler \citep{Goodman_2010}, was used to obtain the best fit of the parameters and their posterior probabilities based on the observed VHE data. This approach is based on Bayesian statistics with a set of parameters (hereafter, $\theta$) for the data (hereafter, $\mathscrsfs{D}$). The \textbf{emcee} sampling algorithm generates multiple MCMC chains (`walkers') that evolve through the parameter space simultaneously at each step. Assuming that the observed data points are independent of each other and each one follows a Gaussian normal distribution, the~likelihood function can be expressed as:
\begin{align}
    \mathscrsfs{L}(\theta|\mathscrsfs{D}) = \prod_{i = 1}^{m}\frac{\exp\left(-\left((\frac{dN}{dE})_{i,obs} - (\frac{dN}{dE})_{i}\right)^{2}/2\sigma_{i,obs}^{2}\right)}{\sqrt{2\pi\sigma_{i,obs}^{2}}}
\end{align}
where $(\frac{dN}{dE})_{i}$ is the model flux, $m$ is the number of data points, and $(\frac{dN}{dE})_{i,obs}$ and $\sigma_{i,obs}$ are the observed flux and the variance, respectively.

Following Baye's theorem, the~posterior probability is given by:
\begin{align}
    P(\theta|\mathscrsfs{D}) \propto \mathscrsfs{L}(\mathscrsfs{D}|\theta)P(\theta). 
\end{align}
Due to the simplicity of the models, we used uninformative (flat) priors ($P(\theta)$) for all parameters with appropriately large~limits.

To determine the best EBL shape that describes the near-infrared (NIR) range, three plausible shapes motivated by the current understanding/constraints of EBL flux density have been explored thoroughly and are listed in Table~\ref{Table: 1}. Since the EBL range implied by the OJ 287 VHE spectrum is very small, we adopted the power-law (PL) shape \citep{Jager1994}. However, the~first EBL peak (Figure \ref{fig:Fig. 1}) resides within the constraint range, and thus, complex models parabola (PB, with~axis $\lambda_{0} = 1.4~\upmu$m) and polynomial (PM) were considered to study the possible curvature information~present.

\begin{table}[H]
    \caption{Functions describing the different EBL~forms.}
    \setlength{\tabcolsep}{4.21mm}

    \begin{tabular}{ c c c c }
    \toprule
    \textbf{Name} & \textbf{Abbreviation} & \textbf{Function} & \textbf{Parameters to Evaluate} \\     
    \midrule
    Power Law & PL & $\phi_{0}(E)^{-\Gamma}$ & $\phi_{0}$, $\Gamma$ \\

    Parabola & PB & $-4h(\lambda - \lambda_{0})^{2} + k$ & $h, k$ \\

    Polynomial & PM & $a\lambda^{3} + b\lambda^{2} + c\lambda + d$ & $a, b, c, d$ \\
    \bottomrule
    \end{tabular}
    \label{Table: 1}
\end{table}

The fitting procedure consists of the following steps: (i) One of the three functional forms was chosen to model the EBL. (ii) We initialized the code with 64 walkers exploring the parameter space in 10,000 steps each, with~a burn-in of 2000 steps for each walker. For~chains to converge we sampled for a large number of steps with different walkers initialized differently and then ran with different random number seeds to create posterior probability inferences that are substantially the same. (iii) Finally, the~posterior probability and the likelihood profile were computed as a function of the model parameters ($\theta$) to estimate the statistical uncertainty associated with the maximum likelihood estimates (MLEs).

To select the EBL model most compatible with the observational data, we used various numerical procedures to obtain the goodness-of-fit and to determine the corresponding best value of the fitting parameters. In~general, the~least chi-squared ($\chi^{2}$) method is sufficient for comparing models, and the~model with the least $\chi^{2}$ value implies that it has a more acceptable fit with the data than the others. However, a reduced chi-squared ($\chi_{red}^{2}$) is used extensively in testing the goodness-of-fit, so we have considered $\chi_{red}^{2}$ for a better interpretation of the result, which is defined as follows:
\begin{align}
    \chi^{2}_{red} = \frac{\chi^{2}_{min}}{N - K},
\end{align}
where $N$ is the number of data points, $K$ is the number of free parameters for the model, and for~Gaussian errors, $\chi^{2}_{min} = -2ln((\mathscrsfs{L}(\theta|\mathscrsfs{D}))_{max})$. When the variance of the measurement error is $\chi_{red}^{2} \approx 1$, this implies that the measurement error is in accord with the error variance. Having the value of $\chi^{2}_{red} \gg 1$ indicates a poor model fit. A~$\chi_{red}^{2} \ll 1$ indicates that the model has over-fit the data are or the variance of the measurement error has been overestimated.
Since the degree of freedom (N-K) is small, $\chi^{2}$ efficiency reduces greatly when comparing more complex models; to make a fair comparison, we considered two advanced information criteria: the Akaike Information Criterion (AIC) by \citet{akaike1973information} and the Bayesian Information Criterion (BIC) by \citet{Schwarz1978}, defined as:
\begin{align}
    AIC = \chi^{2}_{min} + 2K
\end{align}
\begin{align}
    BIC = \chi^{2}_{min} + K~ln(N).
\end{align}

Now, after~considering a reference model (denoted by $R$), for~any model $M$, the difference $\Delta$AIC$= |\text{AIC}_{M} - \text{AIC}_{R}|$ will get us the following conclusions. As explained in \citet{Burnham2004}, (i) if $\Delta$AIC$ \leq 2$, then there is significant support for the $M^{th}$ model, and~$M^{th}$ is probably the proper description of the data, (ii) $4 \leq \Delta\text{AIC} \leq 7$ implies less support for the $M^{th}$ model over the reference model, and, (iii) $\Delta$AIC$\geq 10$ means that the $M^{th}$ model essentially has no support, in~principle, this model is of no use. For~BIC model selection criteria, following the guidelines of \citet{Raftery1995}, if~the difference in the BICs is $0 < \Delta\text{BIC} \leq 2$, this implies weak evidence, for~$2 \leq \Delta\text{BIC} \leq 6$, this implies positive evidence, and~$6 \leq \Delta\text{BIC} \leq 10$ is strong evidence, whereas more than 10 is very strong evidence against the model with a higher~BIC.

\section{Results}\label{Results}
\begin{figure}[H]
    \begin{subfigure}{.5\textwidth}
        \includegraphics[width= \linewidth, height= 6cm]{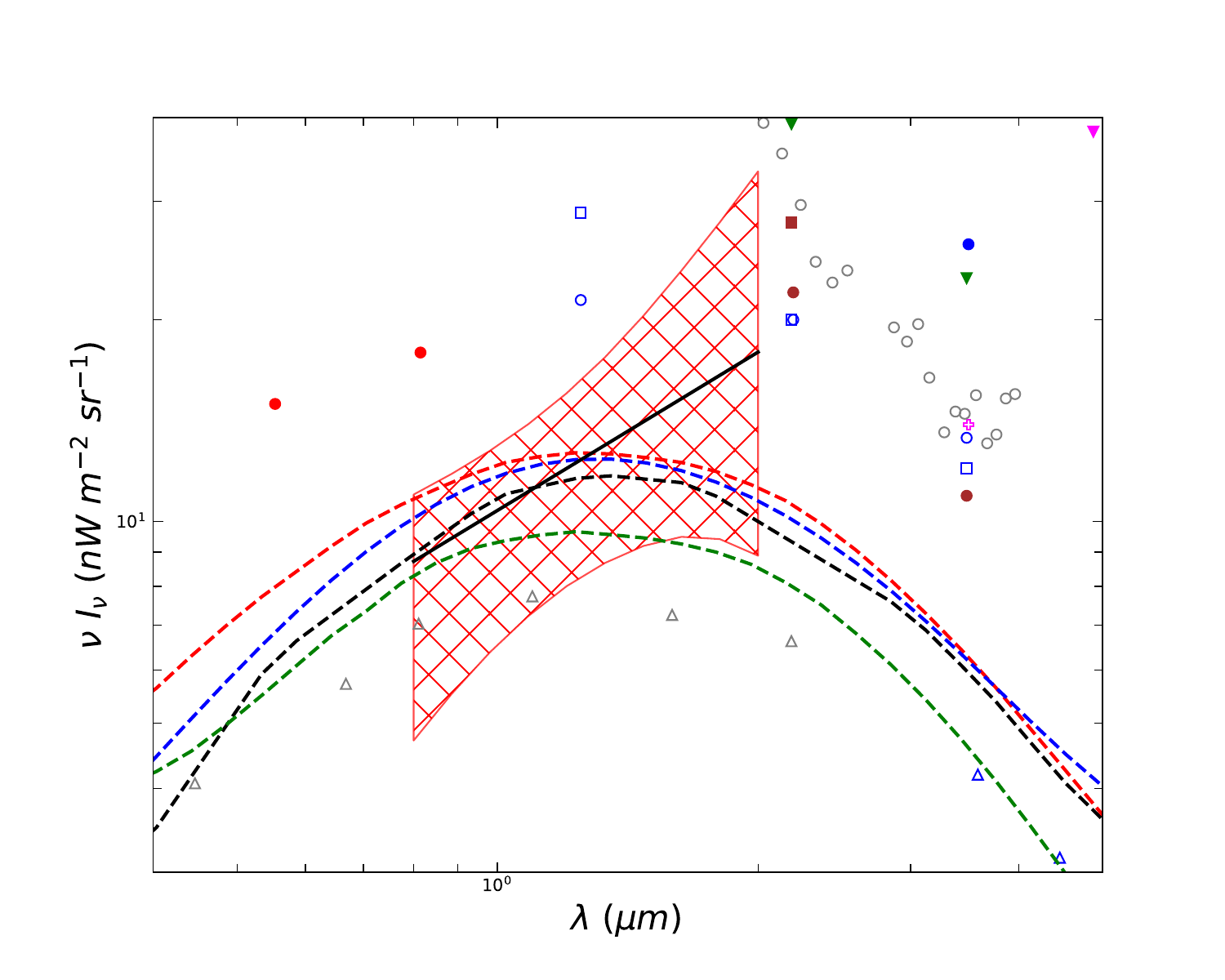}
        \caption{EBL model: PL: (i)}
        \label{fig:3a}
    \end{subfigure}%
    \begin{subfigure}{.5\textwidth}
        \includegraphics[width= \linewidth, height= 6cm]{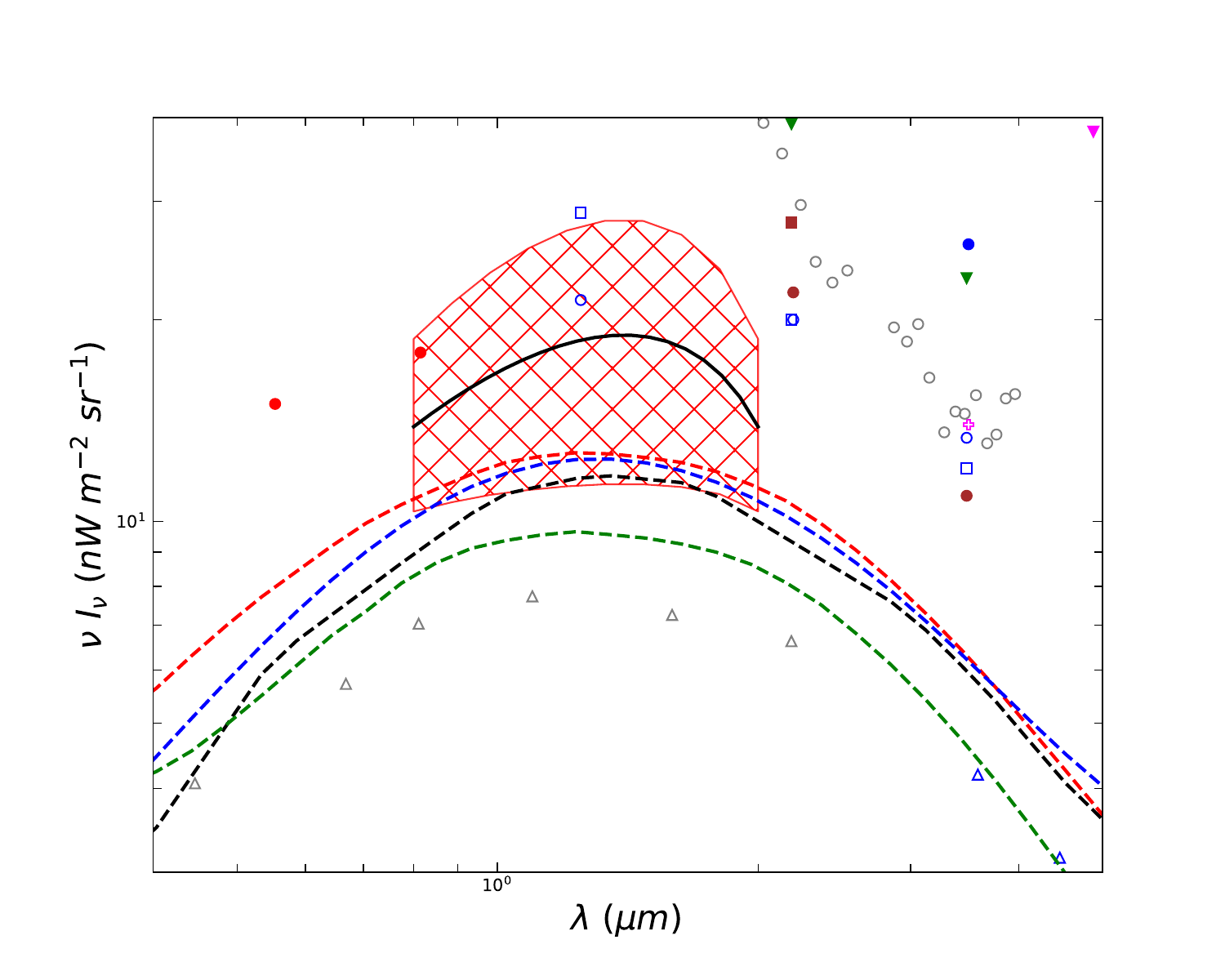}
        \caption{EBL model: PB: (i)}
        \label{fig:3b}
    \end{subfigure}%
    \vskip4ex
    \begin{subfigure}{.5\textwidth}
        \includegraphics[width= \linewidth, height= 6cm]{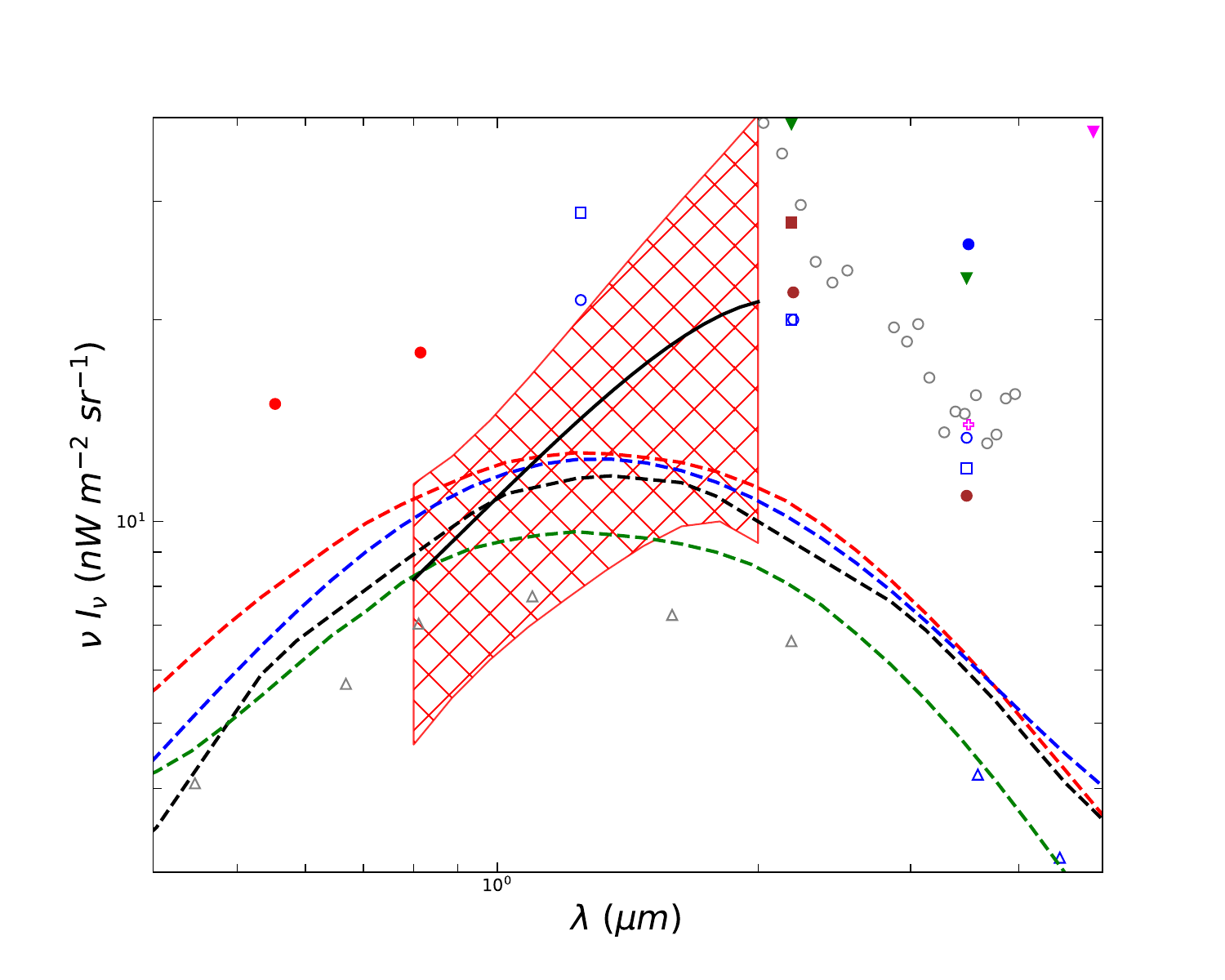}
        \caption{EBL model: PM}
        \label{fig:3c}
    \end{subfigure}%
    \begin{subfigure}{.5\textwidth}
        \includegraphics[width= \linewidth, height= 6cm]{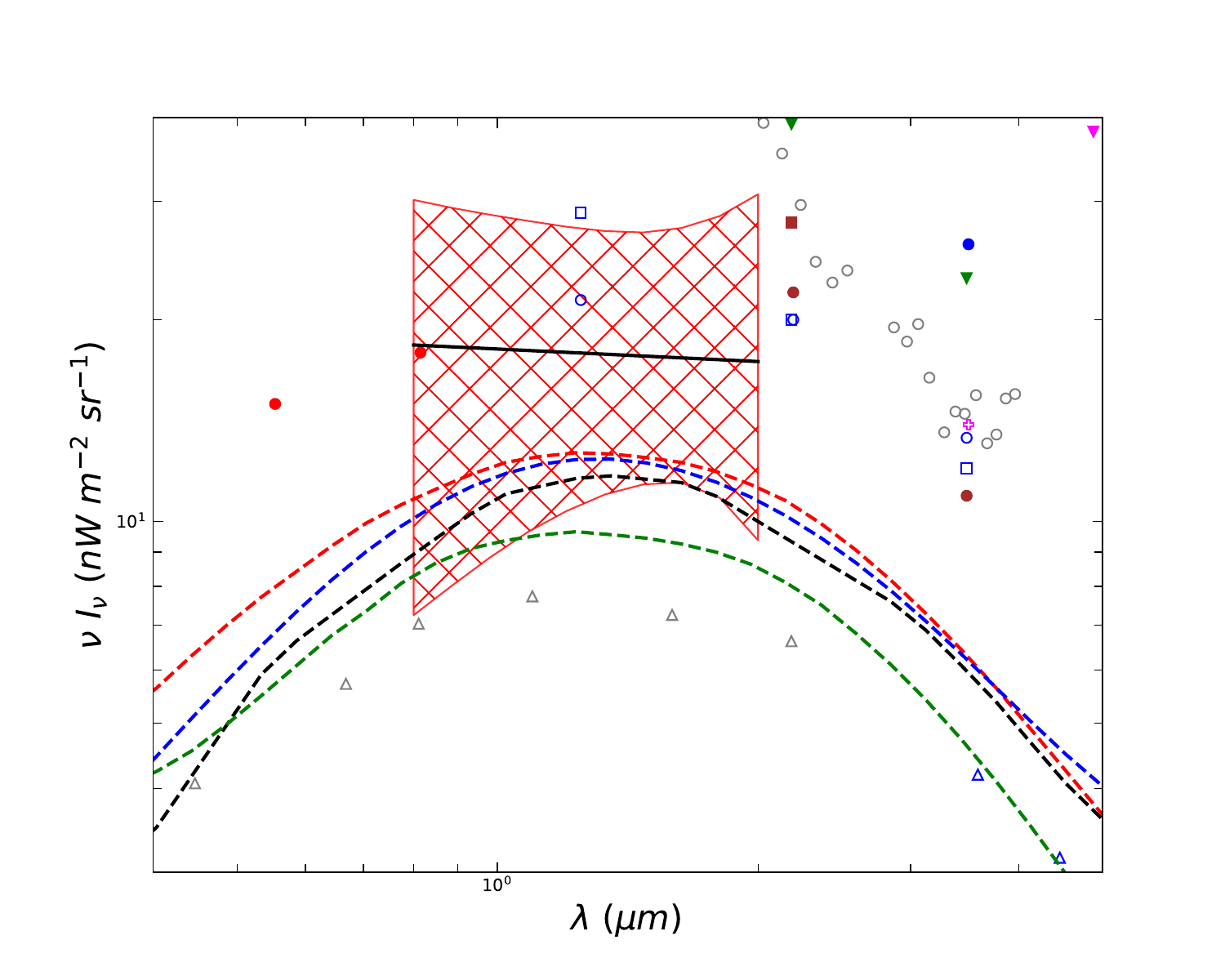}
        \caption{EBL model: PL: (ii)}
        \label{fig:3d}
    \end{subfigure}%
    \vskip4ex
    \centering
    \begin{subfigure}{.5\textwidth}
        \includegraphics[width= \linewidth, height= 6cm]{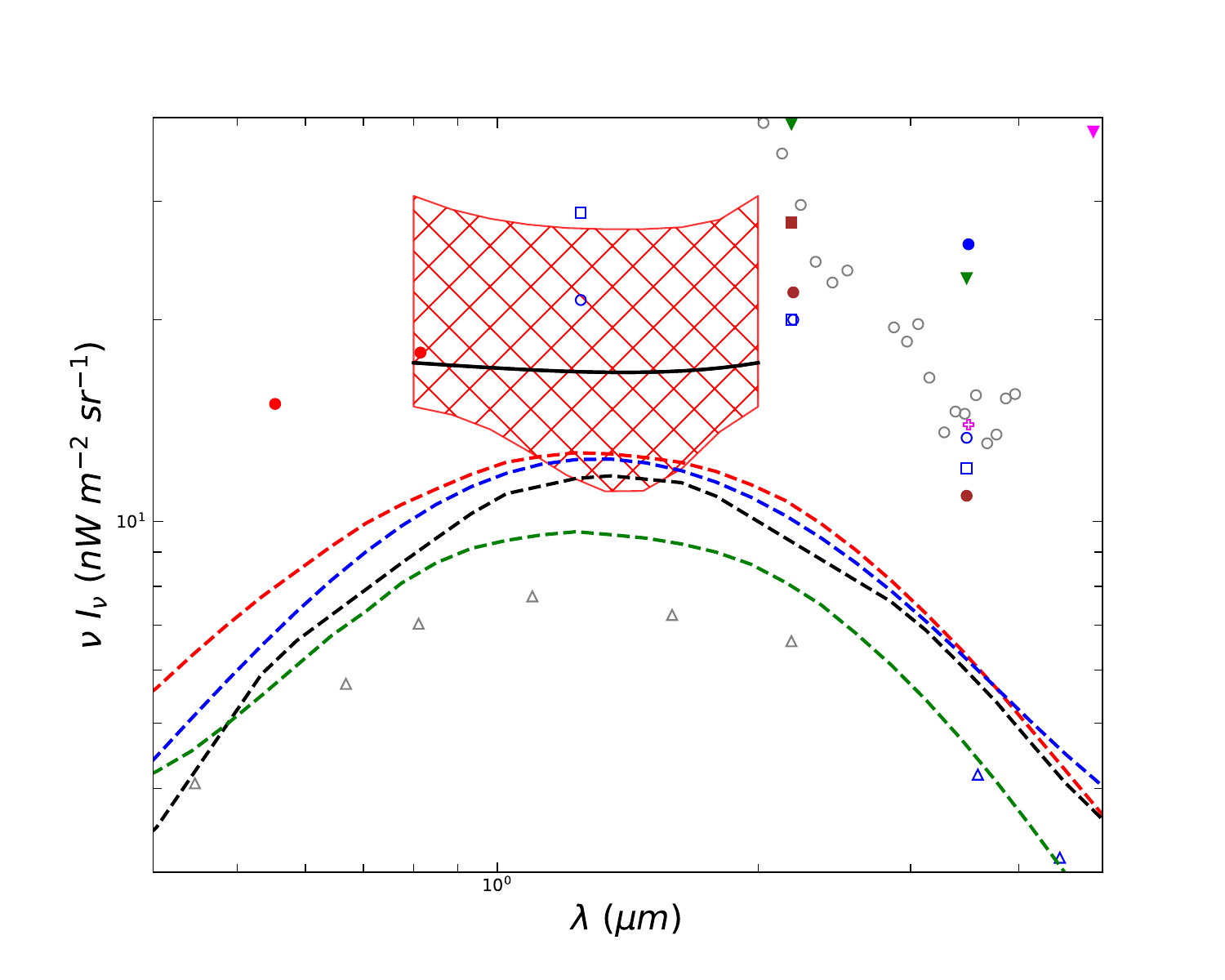}
        \caption{EBL model: PB: (ii)}
        \label{fig:3e}
    \end{subfigure}%
    \begin{subfigure}{.5\textwidth}
        \includegraphics[width= \linewidth, height= 6cm]{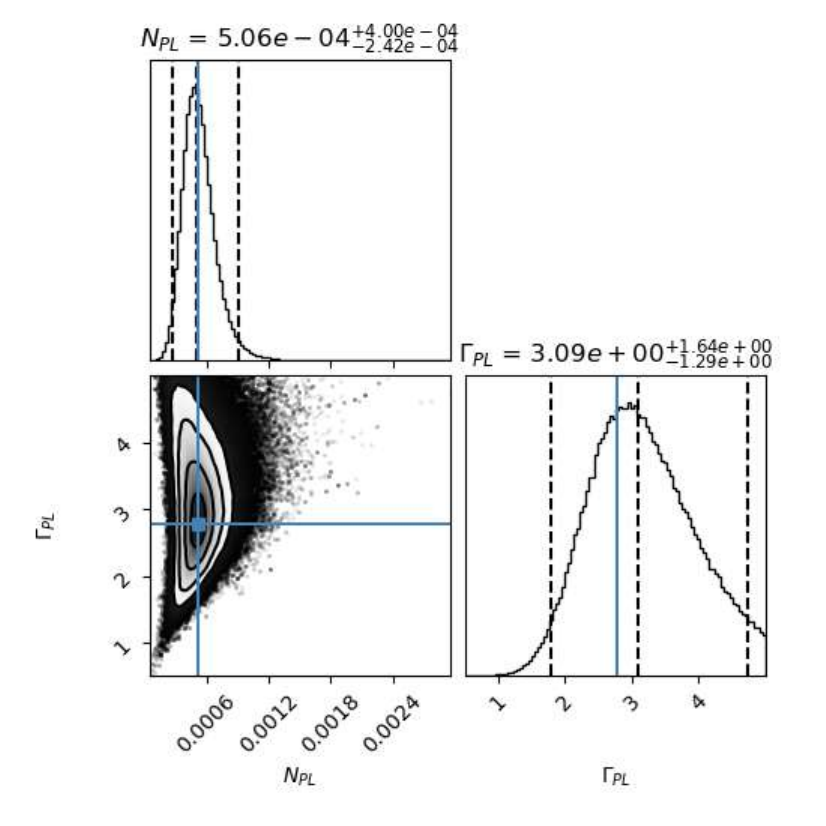}
        \caption{Posterior distribution of PL: (i)}
        \label{fig:3f}
    \end{subfigure}%
    \vspace{+6pt}
    \caption{The best-fit (solid black curves) (\textbf{a}--\textbf{e}) with $68\%$ (1$\sigma$, red-shaded) confidence intervals for the EBL intensity along with the EBL model curves (dashed) by \citet{Franceschini_2008} (black), \citet{Gilmore_2009} (red), \citet{Finke_2010} (blue), and \citet{Kneiske_2010} (green) are plotted for reference. Panel (\textbf{f}) shows the corner plot for the PL model (the best description; see Section \ref{Method}). }
    \label{fig:Fig. 3}
\end{figure}

Figure~\ref{fig:Fig. 3} shows the best-fit EBL curves for different assumed phenomenological models listed in Table~\ref{Table: 1} along with the $1\sigma$ credible range, extracted from the posterior. The~corresponding best-fit parameters are given in Table~\ref{Table: 2}.

The best-fit model parameters are derived following the methodology outlined 
in the previous section. We have three models: power-law (PL), parabola (PB), and~a polynomial (PM). The~respective functional form is given in Table~\ref{Table: 1}.
Of these, the~PM model has the maximum number of parameters (four), while the PL and PB have
two each.
Given this and noting that we have only five VHE data points, there are possibilities
in terms of fixing/freezing a few of the parameters in different models based on our current understanding. The~simplest is the normalization factor. The~analysis is performed by taking two cases into account: (i) Considering that the $\gamma$-ray absorption at $E_{VHE} \sim 100~\text{GeV}$ is negligible, hence fixing the normalization of the intrinsic VHE spectrum to the first VHE observed data at $E_{VHE} = 0.1187~\text{TeV}$, and (ii) keeping the VHE flux normalization factor as a fitting parameter along with the EBL model~parameters.

In Table~\ref{Table: 2} and the best-fit plot (Figure \ref{fig:Fig. 3}), the~$1\sigma$ range is the median, the $68\%$ 
credible intervals (analogous to confidence intervals in frequentist statistics) of the model parameters resulting from the MCMC fits. 
The table also summarizes the parameters that were fixed or free during the model fits and the best-fit criteria:~$\chi^{2}$,~$\chi^{2}_{red}$, AIC, and BIC.
The $1\sigma$ confidence bands for all the EBL models are in good agreement with the lower and upper limits from direct measurements. Since
the data available are very limited, this causes the confidence intervals to be very~wide.

\begin{table}[H]\renewcommand{\arraystretch}{1.2}
    \caption{The median and $68\%$ credible intervals of free parameters resulting from MCMC fits along with the statistics~value.}
    \setlength{\tabcolsep}{2.55mm}

    \begin{tabular}{c c c c c c}
    \toprule
    \centering \textbf{Model: (Case)} & \centering \textbf{Parameters} & \boldmath{$\chi^{2}$} & \boldmath{$\chi^{2}_{red.}$} & \textbf{AIC} & \textbf{BIC} \\     
    \midrule
    PL: (i) & \makecell{$N_{VHE} = Fixed$ \\ $N_{ebl} = 5.1^{+4.0}_{-2.4}\times10^{-4}$ \\ $\Gamma_{ebl} = 3.1_{-1.6}^{+1.3}$} & 0.434 & $0.434/3$ & $-$249.740 & $-$247.938 \\
    \midrule
    PB: (i) & \makecell{$N_{VHE} = Fixed$ \\ $h = 9.4^{+7.2}_{-4.8}\times10^{-1}$ \\ $k = 1.3^{+1.8}_{-0.8}\times10$ } & 1.204  & $1.204/3$ & $-$248.970 & $-$247.168 \\
    \midrule
    PM & \makecell{$N_{VHE} = Fixed$ \\ $a = -0.1^{+1.0}_{-0.8}\times10$ \\ $b = 1.2^{+3.5}_{-3.0}\times10$ \\ $c = -0.6_{-3.2}^{+3.9}\times10$ \\ $d = 0.3^{+1.3}_{-1.1}$ } & 0.431  & $0.431/1$ & $-$247.546 & $-$243.941 \\
    \midrule
    PL: (ii) & \makecell{$N_{VHE} = 2.5_{-1.4}^{+6.8}\times10^{-10}$ \\ $N_{ebl} = 7.9_{-6.3}^{+6.8}\times10^{-4}$ \\ $\Gamma_{ebl} = 2.1_{-0.8}^{+3.1}$} & 0.218 & $0.218/2$ & $-$248.858 & $-$246.154 \\
    \midrule
    PB: (ii) & \makecell{$N_{VHE} = 3.1_{-1.5}^{+4.1}\times10^{-10}$ \\ $h = -2.1_{-7.4}^{+8.7}$ \\ $k = 2.2^{+1.7}_{-1.4}\times10$} & 0.208 & $0.208/2$ & $-$248.867 & $-$246.163 \\
    \bottomrule
    \end{tabular}
    \label{Table: 2}
\end{table}

Secondly, to~study the best possible shape describing the constraint EBL flux density region, we thoroughly examined three shapes (see Table~\ref{Table: 1}).
The PB and PM models have the perk of fitting a wide range of curvature despite that this roughly resembles a power-law-like shape (see Figure~\ref{fig:Fig. 3}).
Using the AIC and BIC analysis, no significant ($\Delta\text{AIC}, ~\Delta\text{BIC}$) difference is observed between the results of the models, and we
concluded that all the EBL models fit the constraint region well. For~the PL model, the~AIC and BIC are smaller than the others, implying that the power-law
is the best description for the constrained EBL flux density region. The~corresponding corner plot is in the bottom-right panel in Figure~\ref{fig:Fig. 3}.

\begin{figure}[H]
  \includegraphics[scale=0.8]{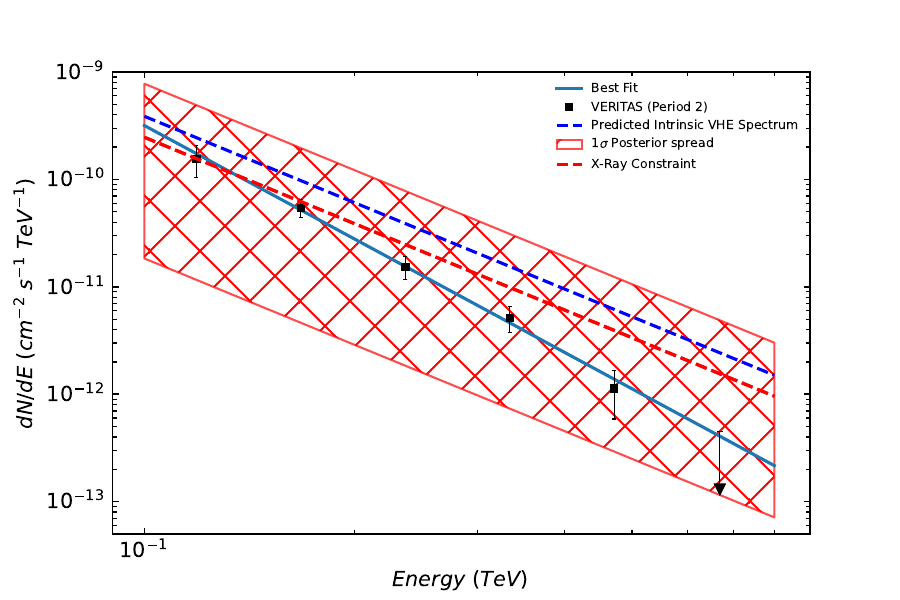}
  \caption{The constructed EBL-corrected VHE spectrum (blue dashed, PL (ii)) along with the $1\sigma$ confidence band (striped red region). The~X-ray spectrum is plotted for reference (normalization scaled to match the VHE data)}
  \label{fig:Fig. 4}
\end{figure}

Thirdly, the~predicted EBL by PL (i) model fitting suggests the increase in the EBL flux density beyond wavelength $\lambda$$\sim$$1.4~\upmu$m, not in agreement with our present knowledge of the EBL. However, it is in between the constraints derived with galaxy counts and the direct measurements (see Figure~\ref{fig:Fig. 3}a). For PL (ii) scenario, it predicts roughly flat EBL density in the range $\lambda$$\sim$(0.8--2)~$\upmu$m, which is in agreement with the concave-like SED of the EBL at around $\lambda$$\sim$$1.4~\upmu$m (see Figure~\ref{fig:Fig. 3}d). The~$1\sigma$ contour lies in between the constraints derived from the galaxy counts and the direct measurements with a peak amplitude at $1.4~\upmu$m of $\nu I_{\nu}= 17.700~n\text{Wm}^{-2}\text{sr}^{-1}$. The~$1\sigma$ confidence band of the intrinsic VHE spectrum for PL (ii) is in agreement with the constrained intrinsic VHE spectrum as for case (i) (see Figure~\ref{fig:Fig. 4}).

\section{Discussion and~Conclusions}

The gamma-ray emission in BL Lacs with the LBL SED is primarily due to IC scattering of
IR photons (e.g., \citep[][]{Kushwaha_2013, Arsioli_2018,2022ApJ...924...57R}) and thus, for such
sources, if detected at VHE, the~softening due to the KN is expected to be considerably less
compared to the HBLs/HSPs that dominate VHE detected sources. This makes these sources an excellent
EBL estimator and can provide a good re-check of the existing models and constraints on the EBL.
However, only three such BL Lacs have been detected at VHE to date, limiting such tests over a
very limited range of EBL energies. One of these is OJ 287, which has exhibited an almost non-variable
VHE activity over a rather extended duration \citep{obrien2017veritas,2017ATel10051....1M}. The~activity was coincident
with the observation of a new jet feature in the parsec-scale jet \citep{2022A&A...658L..10L}
at $\sim$10 pc, indicating the new HBL-like emission component being associated with it and 
supporting the IC-IR origin of gamma-rays. The~extensive
observation at the optical to X-ray spectra provides excellent simultaneous data to explore the EBL. Using
this and the VERITAS  VHE spectrum data of OJ 287 in the range from 100 to 560 GeV,
we extracted the maximum possible and the best functional form explaining the spectral shape of the EBL in the NIR range. Using the AIC and BIC to compare the models, we concluded that the power-law (PL) model remains the best explanation for the constraint EBL range. Meanwhile, the~predicted EBL flux density for all models is in accord with the limits derived by direct measurements (Figure \ref{fig:Fig. 3}). The~assumption we made on the intrinsic VHE spectrum is the up-scattering of soft photons ($\epsilon_{IR}$), and 
its spectral shape is related to the corresponding synchrotron spectrum
i.e., X-ray here. The~derived limit in the NIR region indicates that the intergalactic space is less transparent to $\gamma$-rays than predicted by the integrated light of resolved galaxies, as~has been reported in other~studies.

The other BL objects with LBL SED detected at VHEs are \textit{OT 081} \citep{2016ATel.9267....1M, Fermi-LAT:2021zzm} and {\textit{AP Librae} \citep{2010ATel.2743....1H}.
However, the OT 081 broadband SED associated with the VHE activity is unavailable
to date \citep{Fermi-LAT:2021zzm}, while the AP Librae VHE SED is argued to be due to IC scattering
of CMB radiation from the extended jet \citep{2015MNRAS.454.3229S}, while MeV--GeV (Fermi-LAT)
from the IC of the IR field has been recently\mbox{ observed \citep{2022ApJ...924...57R}}. The~VHE emission
seems to be the continuity of the gamma-ray spectrum from the Fermi-\mbox{LAT \citep{2022ApJ...924...57R}}. However, the~corresponding synchrotron component is missing and, thus, the intrinsic
shape of the VHE spectrum, which is essential in the current approach. Furthermore, being
a very nearby source, for~the energy range of VHE detection and the uncertainty on the VHE
data, the~gamma--gamma opacity is too small to be significantly reflected in the VHE~spectrum.}

Approaches exploiting the Bayesian methodology have been used to obtain the EBL intensity in the optical band by using 259 VHE spectra from the Spectral TeV Extragalactic Catalog (STeVECat) from 56 extragalactic sources with $z > 0.01$. Being a fully Bayesian framework enabled improvement in the typical resolution of gamma-ray measurements of the EBL to around $15 \%$ \citep{2023arXiv230400808G}. Another recent work by \citet{2024JCAP...03..020G} used the Bayesian with Hamiltonian Monte Carlo approach on 65 VHE spectra of 36 AGNs observed by the Imaging Atmospheric Cherenkov Telescopes (IACTs), simultaneously predicting the EBL flux density in the mid-IR region and intrinsic spectral parameters. For~constraining the IR part
of the EBL, the~authors have considered three blackbody emission components at 40 K, 70 K, and~450 K, representing different dust contributions. They found that the 
observation of the $\sim$20 TeV VHE from the nearby HBL Mkn 501 during a flare (HEGRA/1997)
is a crucial input in constraining the far-IR~region.

Estimating the EBL using VHE is still evolving, primarily due to large uncertainties in the VHE data due to the counting statistic and detection of mostly
low luminous nearby sources (HBLs) 
providing input over a very limited energy range of the EBL. However, the~indirect constraints on the NIR by studying the EBL imprints on TeV $\gamma$-rays will continue to improve as more sources over a greater range of redshifts are observed. Next-generation TeV $\gamma$-ray telescopes, such as the \textit{Cherenkov Telescope Array (CTA)} \citep{ACHARYA_2013,2019scta.book.....C}, with~an order of magnitude better sensitivity and much-expanded energy range from $\sim$20 GeV up to $\sim$300 TeV, will be able to detect fainter sources at higher redshift, thereby
significantly expanding the sample of VHE blazars over both the energy and redshift range. The~low-energy part of the gamma-ray spectrum will
allow an unhindered view of the unaffected spectrum, while the high-energy part will reflect the EBL signature and, especially, the expected
peculiar features like flattening, cutoff, etc., (e.g., \citep[][]{2013APh....43..241M}) serving as a strong test of the existing EBL models,
a tighter constraint on the EBL spectrum, as well as a constraint on the contribution from faint galaxies. 
The coverage will allow a thorough reconstruction of the EBL spectrum over the entire UV-to-IR from the 
current existing VHE samples that are constrained primarily to the UV, optical, and mid-IR \citep{2024JCAP...03..020G}, its evolution, and 
thereby, indirectly an improved and better reconstruction of cosmic star formation history to higher redshifts than the
current one (e.g., like in \citep[][]{Fermi_2018}).

\section*{author contributions}
P.K. conceptualized the work and provided feedback on numerical analysis and implementaion. S.Y. did the numerical coding, analysis, plot generation, and wrote most of the paper. All authors have read and agreed to the published version of the manuscript.

\section*{funding}
P.K. acknowledges support from the Department of Science and Technology (DST), the Government of India, through the DST-INSPIRE faculty grant (DST/ INSPIRE/04/2020/002586).

\section*{data availability}
The X-ray data from Swift-XRT and software needed for reduction
are publically available on HEASARC webpage, \url{https://heasarc.gsfc.nasa.gov/} (Archive and Software links). The VHE data is taken from \citep{obrien2017veritas}.

\section*{acknowledgments}
The authors thank the referees for their constructive comments and criticism that improved the manuscript significantly.

\section*{conflicts of interest}
The authors declare no conflicts of~interest. 

\bsp	
\label{lastpage}
\end{document}